%% file: cf.tex
\documentclass{tlp}

\title[Theory and Practice of Logic Programming]{\bf 
    Fast Frequent Querying with Lazy Control Flow Compilation}

\author[R. Tron\c{c}on et al.]
    {Remko Tron\c{c}on\thanks{Remko Tron\c{c}on is supported by the Institute for the Promotion of
Innovation by Science and Technology in Flanders (I.W.T.).},
    Gerda Janssens, Bart Demoen, Henk Vandecasteele \\
     K.U.Leuven -- Department of Computer Science \\
     \email{\{remko,gerda,bmd,henkv\}@cs.kuleuven.be}}
\submitted{15 April 2005 }
\revised{13 October 2005} 
\accepted{13 January 2006}

\newcommand{\na}{-}

\begin{document}


\maketitle

\begin{abstract}
Control flow compilation is a hybrid between classical WAM compilation and 
meta-call, limited to the compilation of non-recursive clause bodies. This approach is used
successfully for the execution of dynamically generated queries in an inductive logic
programming setting (ILP).  Control flow compilation reduces compilation
times up to an order of magnitude, without slowing down execution.  
A lazy variant of control flow compilation is also presented. By compiling
code by need, it removes the overhead of compiling unreached code (a
frequent phenomenon in practical ILP settings), and thus reduces the size 
of the compiled code. Both dynamic compilation approaches have been
implemented and were combined with query packs, an efficient ILP
execution mechanism.  It turns out that locality of data and code is
important for performance.  The
experiments reported in the paper show that lazy control flow compilation
is superior in both artificial and real life settings.
\end{abstract}



\section{Introduction}

In the context of inductive logic programming (ILP), a large number
of queries is generated dynamically, and then run on a large set of examples. 
From
the data mining point of view, such a query is a hypothesis, and by running
the query on the examples one can check how well the hypothesis covers these
examples.  From an implementor's point of view, every example is a Prolog
program, and the queries are just Prolog queries that have to be executed 
against this large number of programs.  Previously, only 2 options were considered: either the queries are meta-called, or
the queries are first compiled to more efficient WAM bytecode after which
this code is executed.  The second option was identified to be the best in
the ILP setting \cite{Blockeel2002}.  Moreover, query packs \cite{Blockeel2002} were 
developed
as a specialized execution mechanism for executing large sets of queries,
improving execution time up to a factor 100. 
Adapting meta-call to handle these query packs is difficult and inefficient,
and therefore compilation is needed \cite{Blockeel2002,Demoen1999}.

However, experiments indicate that even though compilation improves the
total execution time of queries, the compilation time often dominates the 
total time of an ILP run.  This raises the question of what can be
done to decrease the compilation time, or, in other words, how we can
simplify the compilation step. \emph{Control flow compilation} realizes
this by only compiling the control flow of a query (which deals for example
with the selection of branches in a disjunction), and by using special
meta-call like WAM instructions for calling predicates, thus omitting the
need to do the complex step of setting up arguments to calls.

In this paper we propose control flow compilation as a novel fast compilation 
scheme for queries. This compilation
scheme does not affect query execution performance, and moreover
can be performed in a lazy (or `just-in-time') fashion. The contributions
of this paper are:
\begin{itemize}
\item Control Flow Compilation, which is a hybrid approach between
meta-calling and classical compilation.  This scheme incorporates
the best of both worlds: it has the fast execution times of compiled
code, without needing the expensive compilation step (which is a
dominating factor in practical ILP settings).
\item Lazy control flow compilation, which is a Just-In-Time (JIT) version 
of the control flow compilation scheme. Unreachable parts of the code are not
compiled; this reduces both the compilation time and the code size.
\item An evaluation of integrating (lazy) control flow compilation in 
a practical ILP system. A fast light-weight compiler for (lazy) control flow
compilation was implemented in the hipP system, a Prolog system with specific 
support for ILP (such as query packs). 
\end{itemize}
The topic of this paper was already introduced in 
\cite{Troncon2003c,Troncon2004},  discussing an experimental implementation
and some preliminary results on this and related techniques. \\

The organization of the paper is as follows: in
Section~\ref{sec:motivation}, we briefly sketch the ILP context, the
setting that motivated our work.
In Section~\ref{sec:cfcomp}, control flow compilation is introduced
and evaluated on both artificial and real life benchmarks. A lazy
variant of this scheme is introduced in Section~\ref{sec:lazyqueries}.
Both approaches are then adapted to a practical ILP setting, by
extending them to the query packs execution mechanism in 
Section~\ref{sec:lazypacks}. This extension is again evaluated on 
real life ILP benchmarks. Section~\ref{sec:memory} discusses
memory management issues of the approaches described in this paper.
Finally, conclusions and future work are given in
Section~\ref{sec:conclusion}.

We assume the reader is familiar with the WAM~\cite{Warren1983,Ait-Kaci1990}.

\section{Context: Dynamically Generated Queries in ILP} \label{sec:motivation}
We start by sketching the role of dynamically generated queries and
their execution in the ILP setting. The goal of ILP is to find 
a theory that best explains a large set of data (or examples). 
In the ILP setting at hand, each example is a logic program, and 
the theory is represented as a set of logical
queries. 
The ILP algorithm finds good queries by using generate-and-test. In the first
step, it uses its own specific approach to generate queries, which are
then evaluated in the second step. Based on the failure or success of these queries, 
only the ones with the `best' results are kept and are extended (by adding literals).
Which queries are best depends on the ILP algorithm: for example, in
the case of classification, the information gain can be used as a
criterion, whereas in the case of regression, the reduction of
variance is often used.  The extended queries are in turn tested on a
set of examples, and this process continues until a satisfactory query
(or set of queries) describing the examples has been found.
We will focus on the efficiency of the second step.

At each iteration of the algorithm, a set of queries is executed against a 
large set of logic programs
(the examples). Since these queries are the result of adding 
different literals at the end of another query, the queries in this set have 
a lot of common prefixes. To avoid repeating the common parts by executing
each query separately, the set of queries can be transformed into a special kind
of disjunction: a \emph{query pack} \cite{Blockeel2002}. 
For example, the set of queries
{\small
\begin{verbatim}
        ?- a, b, c, d.
        ?- a, b, c, e.
        ?- a, b, f, g.
\end{verbatim}}
\noindent 
is transformed into the query
{\small
\begin{verbatim}
        ?- a, b, ( (c,(d;e)) ; f,g ).
\end{verbatim}}
\noindent
by applying left factoring on the initial set of queries.
However, because only the success of a query on an example is 
relevant to the ILP algorithm, the 
normal Prolog disjunction might still cause too much backtracking. So,
for efficiency reasons the \texttt{';'/2} is given a different 
procedural
behavior in query packs: it cuts away branches from the disjunction as 
soon as they succeed. For this paper, it is sufficient to know that the
semantics of query packs is very close to the one of normal
disjunctions.  
Since each query pack is run on a large set of
examples, a query pack is first compiled, and the compiled code is
executed on the examples. This compiled code makes use of special
WAM instructions for the query pack execution mechanism. More details
can be found in \cite{Blockeel2002}.

Our paper proposes a new compilation approach for dynamically generated
queries. It is independent of the actual query optimization schemes used.
Moreover, it is useful for other Prolog engines as well, and is independent 
of the engine-specific support for ILP. After discussing how to use the approach to
compile queries in general, we will apply it to the specific case of query
pack compilation.

\section{Control Flow Compilation} \label{sec:cfcomp}

\subsection{Technology} \label{sec:cfcomptech}

To execute a dynamically generated query, we can either meta-call it, or we
can transform it into a non-recursive clause (by taking the query as the body and adding a 
head, e.g. \texttt{query/0}), and run the compiled clause.
Executing this compiled version of the query instead of meta-calling it results in
considerable speedups. Moreover, in order to benefit from an efficient
execution mechanism such as query packs, queries have to be compiled into 
the special WAM instructions. However, compilation of a query can be a costly
task; in the ILP setting, compilation of a query can take as
much time as its execution on all examples. 
This motivated the preliminary study of alternatives for compile \& run in
\cite{Troncon2003c}. The most interesting alternative turned out
to be \emph{control flow
compilation}, which is a hybrid between meta-calling and compiling a query.
In this section, we introduce control flow compilation for queries whose
bodies consist of conjunctions and disjunctions, and explain this
approach in terms of the familiar WAM instructions. 
The experiments confirm the potential of control flow compilation.
This scheme will be
extended to control flow compilation for ILP query packs in 
Section~\ref{sec:lazypacks}. \\

The essential difference between classical compilation and control flow
compilation is the sequence of instructions generated for setting up and
calling a goal. Instead of generating the usual WAM \texttt{put} and
\texttt{call} instructions, control flow compilation generates one new \texttt{cf\_call}
instruction, whose argument points to a heap data structure (the goal) that
is meta-called.  Hence, control flow code only contains the control flow
instructions (\texttt{try}, \texttt{retry}, \ldots) and \texttt{cf\_call}
(and \texttt{cf\_deallex}\footnote{\texttt{cf\_deallex} is the instruction
obtained by merging \texttt{deallocate} and \texttt{cf\_execute}.}) instructions. 

For example, control flow compiling the query
{\small
\begin{center}
\texttt{query :- a(X,Y), ( b(Y,Z) ; c(Y,Z), d(Z,U); e(a,Y) ).}
\end{center}}
results in the code in the left part of Figure~\ref{fig:cfcomp}.
Note that, because queries are dynamically generated by the ILP system,
the query itself is a term on the heap, and we
use \texttt{\&a(X,Y)} to represent the pointer to its subterm
\texttt{a(X,Y)}.
On the right  of Figure~\ref{fig:cfcomp} is the classical compiled 
code for the same query\footnote{The instruction set used in this example
is based on the XSB implementation:
the suffixes {\em tvar} and {\em pval} correspond to the WAM temporary
and permanent variables respectively.}.
\begin{figure}
\input{fig_cfvscr}
\caption{Control flow compiled code vs. classical compiled code.}
\label{fig:cfcomp}
\end{figure}
Before calling each goal, the compiled code first sets up the
arguments to the goal, whereas the control flow compiled code uses a
reference to the subterm of the query to indicate the goal that is
called. 
The most important aspect is that the control flow code saves emulator
cycles, because it contains no instructions related to the arguments
of the goals that are called.
Moreover, the absence of these kinds of instructions has other 
positive consequences: (1) it makes the expensive (non-linear) argument register 
allocation step unnecessary, saving compilation time, and (2) it makes it
easy to incrementally add new code to existing parts of code.
The latter is
very interesting because it makes introducing laziness in the compilation 
process possible, as explained in Section~\ref{sec:lazyqueries}. \\

Contrary to compiled code, control flow code cannot exist on its own, since
it contains external references to terms on the heap. Therefore, an
implementation must take the following garbage collection issues into 
consideration:
(1) the terms of the query have to be kept alive as long
as the control flow compiled code can be executed; (2) when these terms are
moved to another place in memory, the
references in the code must be adapted as well. 
Moreover, executing general clauses requires that new variables be 
created prior to executing the body, whereas this is not necessary 
for queries: because queries will never be called recursively, the 
variables already existing on the heap can be used. 

\begin{figure}
\include{fig_builtins}
\caption{Built-in inlining for (a(X,Y), X $<$ Y)} \label{fig:built-ins}
\end{figure}
To speed up execution, the classical compilation scheme typically inlines 
smaller predicates (such as tests) using dedicated instructions implemented 
in the system. This is illustrated by the first column of 
Figure~\ref{fig:built-ins}: the WAM compiler initializes the argument registers,
and instead of calling a (WAM-compiled) \texttt{'<'/2} predicate, it
emits a built-in instruction to do the test. 
Since control flow compilation also emits WAM bytecode, the same built-ins
can be used for control flow compiled code as for classical compiled code.
These built-in instructions typically use argument registers for
their arguments, so the compiler just needs to emit extra instructions to move
data structures on the heap into the correct argument registers. These
are illustrated in the third column of Figure~\ref{fig:built-ins}, 
where the \texttt{putarg} instructions move references to data structures
on the heap into the relevant argument registers for the built-in
instruction. Alternatively, the extra emulator cycles needed for filling
the argument registers can be skipped by defining special versions of
each built-in that, instead of
argument registers,  have references to the heap as their parameters, such 
as \texttt{cf\_smaller} in Figure~\ref{fig:built-ins}.

\subsection{Evaluation}
For evaluating our approach, we added support for control flow code to the
hipP system \cite{hipp}, an efficient WAM based Prolog system written in C,
and descendant of ilProlog.
A separate control flow compiler for queries was implemented in this system.
This compiler was written in Prolog, as is the case for the existing
classical compiler.  For the built-in predicates that are frequently used in
ILP applications (e.g. \texttt{'<'/2}, \texttt{'>'/2}, \texttt{'='/2},
\verb+'\='/2+, \ldots), we implemented special control flow instructions 
(such
as \texttt{cf\_smaller} from Figure~\ref{fig:built-ins}), and these built-ins 
are inlined by the
control flow compiler.  The heap garbage collector of hipP was modified to
support control flow compiled code. More details about the memory management
will be discussed in Section~\ref{sec:memory}.

All experiments were run on a Pentium III 1.1 GHz with 2
GB main memory running Linux, with a minimum of applications running.

Two kinds of experiments are discussed: the benchmarks in Table 
\ref{tab:cfcomp_disj} show the potential gain in an artificial
setting, whereas the results in Table~\ref{tab:aceconj} are obtained
from a real world application.

\begin{table}
\input{tab_artif}
\caption{Experiments for artificial disjunctions.}
\label{tab:cfcomp_disj}
\end{table}

For each artificial experiment, a query was generated with the following
parameters:
\begin{itemize}
\item {\em G}: the number of goals in a branch,
\item {\em B}: the branching factor in a disjunction,
\item {\em D}: the nesting depth of disjunctions.
\end{itemize}
%
For example, for the values $G=2$, $B=3$ and $D=1$, the following query 
is generated:
{\small
\begin{verbatim}
        ?- a(A,B,C), a(C,D,E), ( a(E,F,G), a(G,H,I)
                               ; a(E,J,K), a(K,L,M)
                               ; a(E,N,O), a(O,P,Q) ).
\end{verbatim}}
\noindent
For $G=1$, $B=2$ and $D=2$, the generated query has nested disjunctions: 
{\small
\begin{verbatim}
        ?- a(A,B,C), ( a(C,D,E), ( a(E,F,G) ; a(E,H,I) ) 
                     ; a(C,J,K), ( a(K,L,M) ; a(K,N,O) ) ).
\end{verbatim}}
\noindent
The definition of \texttt{a/3} was taken to be  \texttt{a(\_,\_,\_)} to 
minimize the time spent outside of the query execution.
For each generated query, the average compile and run time of the query was 
measured over a significant number of runs. 
We report on the following three 
alternatives:
\begin{itemize}
\item {\em Control Flow}: The query is compiled using the control flow approach 
      before it is executed.
\item {\em Compile \& Run}: The query is compiled using the classical WAM 
      compilation before it is executed.
\item  {\em  Meta-call}:  the query is  meta-called (no compilation at all).
\end{itemize}

The control flow compilation is clearly better than compile \& run:
the compilation times are improved by one order of magnitude, while
the execution times are also improved. The compilation 
in the control flow approach is much faster because it does not need to perform
expensive tasks such as assigning variables to environment slots. The
better execution times are explained by the fact that only one
emulation cycle per call is needed as no arguments have to be put in
registers.  Doubling the {\em G} parameter more or less doubles the
timings. For larger queries, namely for $G=10$, $B=10$, $D=4$, and 
for $G=5$, $B=5$, $D=6$, control flow 
compilation becomes up to a factor 16 faster than compile \& run.
If the query is executed a sufficient number of times, meta-call is outperformed
by control flow compilation (e.g. for $G=5$, $B=5$, $D=4$, this number is
15). 
Since in ILP, each query is run on thousands of examples,
these results are very promising. \\

\begin{table}
\input{tab_aceconj}
\caption{Experiments for conjunctions from a real world application.}
\label{tab:aceconj}
\end{table}
The real world experiment consists in running the \textsc{Tilde} algorithm 
\cite{Blockeel1998} from the ILP system ACE/hipP \cite{ace} on three well-known
datasets from the ILP community: Mutagenesis \cite{Srinivasan96:jrnl}, Bongard
\cite{DeRaedt95-ALT:proc} and Carcinogenesis \cite{Srinivasan99:proc}.
During the execution of \textsc{Tilde}, queries are subsequently generated, 
and every query needs to be run on a subset of the examples. These
queries consist only of conjunctions, and every query is executed separately
on the examples.
Table~\ref{tab:aceconj} compares the compilation time and
execution time for all queries in the control flow compilation approach
with the corresponding times of the compile \& run and the meta-call
approach. For each dataset, the total number of queries and the average
number of runs per query is also given.

In the \textsc{Tilde} runs, control flow compilation gains a factor 5 to 8
over usual compilation. For all datasets, control flow compiled
code also outperforms both the classical compiled code and the meta-called
queries. Meta-call is slower than control flow compiled code because of the
extra emulator cycles spent in testing the incoming goal upon each call.
Additionally, specialized variants of \texttt{cf\_call} are used for
calling goals with arities smaller than 4 (which are the most frequent in
practical ILP applications).

We conclude that control flow compilation is the fastest approach for 
executing the queries on these datasets. The main reason for this is that
the share of query compilation in the total execution time of the ILP 
algorithm is reduced significantly. Moreover, control flow compiled
code contains less instructions, and as such saves emulator cycles as
well. \\

The results are more pronounced for the artificial benchmarks than for
the \textsc{Tilde} ones for several reasons.
The artificial queries are longer than the typical \textsc{Tilde}
queries; making the artificial queries shorter makes the timings
unreliable.
During the artificial benchmarks, the time spent in the called goals
is very small (only \texttt{proceed}), whereas in the \textsc{Tilde}
experiments much more time is spent in the predicates, and as such the
effect of control flow on the {\em exec} timing decreases.

\subsection{Conclusion}
The main goal of control flow compilation was to reduce high
compilation times, without slowing down execution itself.
Our experiments prove that control flow compilation achieves this goal: 
compilation times are reduced by an order of magnitude, while the execution
becomes even slightly faster. Moreover, the new compilation scheme is 
flexible, and allows for extensions such as lazy compilation, as 
will be discussed in Section~\ref{sec:lazyqueries}.

\section{Lazy Control Flow Compilation} \label{sec:lazyqueries}

\subsection{Technology}
In practical ILP applications, it is observed that large parts of the
queries generated by the query generation process are never executed.
Hence, unnecessary time is spent in compiling this unreachable code. With
a lazy compilation scheme which only compiles code when it is actually
reached, this redundancy can be removed. Control flow compilation is
particularly suited for this dynamic kind of code, since existing
compiled code can be extended without needing to alter the 
latter because of e.g. argument register allocation (as is the case
with classical compilation). In this section, we will extend the
control flow compilation scheme to yield a lazy variant.

In \cite{jit:acmcs2003}, \emph{lazy compilation} is identified as a kind of 
\emph{just-in-time} (JIT) compilation or \emph{dynamic compilation}, which is 
characterized as translation which occurs after a program begins execution. 
Our lazy variant implicitly calls the control flow compiler when
execution reaches a part of the query that is not yet
compiled. We restrict the
discussion in this section to queries with conjunctions and disjunctions; the
extension to query packs is presented in Section~\ref{sec:lazypacks}. \\

As with normal control flow compilation, the query is represented by a
term on the heap. We introduce a new WAM instruction \texttt{lazy\_compile},
whose argument is a pointer to the term on the heap that needs
compiling when execution reaches this instruction.

Consider the query q :- a(X,Y), b(Y,Z).
The initial lazy compiled version of q is
{\small
\begin{verbatim}
        allocate 2
        lazy_compile &(a(X,Y),b(Y,Z))
\end{verbatim}}
\noindent
The  \texttt{lazy\_compile} instruction points to a conjunction:
its execution replaces itself by the compiled code for  the first
conjunct, namely a \texttt{cf\_call}, and adds
for the second conjunct another \texttt{lazy\_compile} instruction, resulting in:
{\small
\begin{verbatim}
        allocate 2
        cf_call &a(X,Y)
        lazy_compile &b(Y,Z)
\end{verbatim}}
\noindent
The execution continues with the newly generated \texttt{cf\_call}
instruction as is expected. After the next execution of
\texttt{lazy\_compile}, 
the compiled code is equal to code generated without laziness:
{\small
\begin{verbatim}
        allocate 2
        cf_call &a(X,Y)
        cf_deallex &b(Y,Z)
\end{verbatim}}
\noindent
Note that lazy compilation overwrites the \texttt{lazy\_compile}
instruction with a \texttt{cf\_} instruction, and that once we have
executed the query for the first time completely, the resulting code
is the same as the code produced by non-lazy control flow 
compilation. \\

Now, consider the lazy compilation of the query from Figure~\ref{fig:cfcomp}:
{\small
\begin{center}
\texttt{q :- a(X,Y), ( b(Y,Z) ; c(Y,Z), d(Z,U); e(a,Y) ).}
\end{center}}
\noindent
Initially, the code is
{\small
\begin{verbatim}
        allocate 2
        lazy_compile &(a(X,Y),(b(Y,Z);c(Y,Z),d(Z,U);e(a,Y)))
\end{verbatim}}
\noindent
The \texttt{lazy\_compile} changes the code to:
{\small
\begin{verbatim}
        allocate 2
        cf_call &a(X,Y)
        lazy_compile &(b(Y,Z);c(Y,Z),d(Z,U);e(a,Y))
\end{verbatim}}
\noindent
Now, \texttt{lazy\_compile} will compile a disjunction. Where normal
(control flow) compilation would generate a \texttt{trymeorelse} 
instruction, we generate a lazy variant for it. The 
\texttt{lazy\_trymeorelse} instruction has as its argument the
second part of the disjunction, which will be compiled upon failure of
the first branch. The instruction is immediately followed by the
code of the first branch, which is initially again a \texttt{lazy\_compile}:
{\small
\begin{verbatim}
        allocate 2
        cf_call &a(X,Y)
        lazy_trymeorelse &(c(Y,Z),d(Z,U);e(a,Y))
        lazy_compile &b(Y,Z)
\end{verbatim}}
\noindent
Execution continues with the \texttt{lazy\_trymeorelse}: a special
choice point is created such that on backtracking the remaining
branches of the disjunction will be compiled in a lazy way. To achieve
this, the failure continuation of the choice point is set to a new
\texttt{lazy\_disj\_compile} instruction, which behaves similarly to
\texttt{lazy\_compile}. Then, execution continues with the first
branch:
{\small
\begin{verbatim}
        allocate 2
        cf_call &a(X,Y)
        lazy_trymeorelse &(c(Y,Z),d(Z,U);e(a,Y))
        cf_deallex &b(Y,Z)
\end{verbatim}}
\noindent
Upon backtracking to the special choice point created in 
\texttt{lazy\_trymeorelse}, the \texttt{lazy\_disj\_compile} instruction
resumes compilation, and replaces the corresponding \texttt{lazy\_trymeorelse} by a
 \texttt{trymeorelse} instruction 
with the address of the code to be generated as argument: 
{\small
\begin{verbatim}
        allocate 2
        cf_call &a(X,Y)
        trymeorelse L1 
        cf_deallex &b(Y,Z)
    L1: lazy_retrymeorelse &(e(a,Y))
        lazy_compile &(c(Y,Z),d(Z,U))
\end{verbatim}}
\noindent
Here, \texttt{lazy\_retrymeorelse} -- the lazy variant of 
\texttt{retrymeorelse} -- behaves similar to \texttt{lazy\_trymeorelse}, but
instead of creating a special choice point, it alters the existing
choice point. It is immediately followed by the code of the next part
of the disjunction, which after execution looks as follows:
{\small
\begin{verbatim}
        allocate 2
        cf_call &a(X,Y)
        trymeorelse L1 
        cf_deallex &b(Y,Z)
    L1: lazy_retrymeorelse  &(e(a,Y))
        cf_call &c(Y,Z)
        cf_deallex &d(Z,U)
\end{verbatim}}
\noindent
Upon backtracking, \texttt{lazy\_retrymorelse} is overwritten, and
a \texttt{trustmeorelse} is generated for the last branch of the
disjunction, followed by a \texttt{lazy\_compile} for this branch:
{\small
\begin{verbatim}
        allocate 2 
        cf_call &a(X,Y)         
        trymeorelse L1            
        cf_deallex &b(Y,Z)
    L1: retrymeorelse L2
        cf_call &c(Y,Z)
        cf_deallex &d(Z,U)
    L2: trustmeorelsefail 
        lazy_compile &e(a,Y)
\end{verbatim}}
\noindent
After the execution of the last branch, we end up with the full control 
flow code. \\

The lazy compilation as we described proceeds from goal to goal: when the
JIT compiler is called, it compiles exactly 1 goal, and then resumes
execution. Other granularities have been implemented and evaluated as well (see
Table \ref{tab:lazy_disj}):
\begin{itemize}
\item \emph{Per conjunction:} All the goals in a conjunction are compiled at 
      once. This avoids 
      frequent switching between the compiler and the execution by compiling
      bigger chunks.
\item \emph{Per disjunction:} All the branches of a disjunction
are compiled up to the point where a new disjunction
occurs. This approach is reasonable from an ILP viewpoint: the branches of
a disjunction represent different queries, and since the success of each query
is recorded, all branches will be tried (and thus compiled) eventually.
\end{itemize}

Besides the overhead of switching between compilation and execution,
these approaches might also generate different code depending on the
execution itself. When a goal inside a disjunction fails, the next
branch of the conjunction is executed, and newly compiled code is inserted
at the end of the existing code. When in a later stage the same goal succeeds,
the rest of the branch is compiled and added to the end of the code, and a
jump to the new code is generated. These jumps cost extra
emulator cycles and decrease locality of the code.  Lazy
compilation per goal can in the worst case have as many jumps as there
are goals in the disjunctions. Compiling per conjunction can have as
many jumps as there are disjunctions. If a disjunction is completely
compiled in one step, each branch of the disjunction ends in a
jump to the next disjunction. \\

Just as for control flow compilation, special control flow
instructions for built-in predicates can be used in
the lazy variant as well. Care must be taken though: typically, specialized
built-ins are emitted depending on the type of arguments (e.g. specialized
built-ins for unifying arguments with integers); however, 
as compilation is now interleaved with execution, arguments of a goal
might have been bound after starting the execution of the query, which could
make the emitted built-in overly specialized, thus generating code that
becomes erroneous after backtracking or when run on another example.
The compiler therefore shouldn't emit specialized built-ins depending on
the instantiation and/or type of the arguments, or it should keep track of
the state of the goal arguments in the original query. In our
implementation, we chose for the former approach. \\

Finally, note that this lazy control flow compilation can be used to
exploit the incremental nature of a query generation process such as the one
from ILP. Suppose that queries are constructed with an open end, and that the
compiler generates a \texttt{lazy\_compile} instruction for such open ends;
these open ends can be instantiated by a later query generation
phase, such that when execution reaches the \texttt{lazy\_compile}
instruction, the new part of the query will be compiled and added to the
existing code. This avoids the need to recompile the complete query, when
only a small part of it changed. However, as experiments will show, control
flow compilation times are relatively very low, such that the
incremental compilation approach  would not
yield any significant speedups in total query evaluation time with
respect to the use of (lazy) control flow compilation.

\subsection{Evaluation} \label{sec:lazyeval}

In the first experiment, we will measure the overhead of the new lazy 
compilation scheme. The artificial queries from Table~\ref{tab:lazy_disj}
have no unreachable parts, and as such provide a worst case for lazy
compilation overhead. In practical applications, we
expect the queries to have unreachable parts, and so the total overhead of
the lazy compilation scheme will be compensated by the smaller compilation
time. The experiments of Table~\ref{tab:lazy_disj} use only the first
two benchmarks from Table~\ref{tab:cfcomp_disj}. The other
benchmarks of Table~\ref{tab:cfcomp_disj} yield similar results. Timings 
(in milliseconds) are given for the different settings of lazy 
compilation. The timings report the time needed for one execution of the
query, thus including the time of its lazy compilation. These timings
are then compared with the time of performing non-lazy control flow 
compilation of the query and executing it once\footnote{Note that
these timings are slightly higher than the sum of \emph{Comp.} 
and \emph{Exec.} in Table~\ref{tab:cfcomp_disj}.  This is due to the 
fact that both experiments are run in different circumstances with
different locality.}. Lazy compilation per goal clearly has a substantial
overhead, whereas the other settings have a small overhead.
We also measured the  execution times for the three lazy alternatives
once they are compiled: they were all equal, and are therefore not
included in the table.
\begin{table}
\input{tab_lazyartif}
\caption{Lazy compilation for several kinds of disjunctions.}
\label{tab:lazy_disj}
\end{table}

The main message here is that the introduction of laziness in the control
flow compilation does not degrade performance much, and that it opens
perspectives for query packs compilation: (1) lazy compilation is fast; (2)
in real life benchmarks, some branches will never be compiled
due to failure of goals, whereas in our artificial setting all goals in the
queries succeed.

\section{Lazy Control Flow Compilation for Query Packs}
\label{sec:lazypacks}
\subsection{Technology}
So far, we restricted our (lazy) control flow compilation approach to
queries containing conjunctions and `ordinary' disjunctions. However, the
main motivation for this work was optimizing the execution of 
\emph{query packs} \cite{Blockeel2002}. These query packs represent a
set of (similar) queries which are to be executed, laid out in a
disjunction. The semantics of this query pack disjunction is implemented by 
dedicated WAM instructions \cite{Blockeel2002}, as explained in
Section \ref{sec:motivation}. These instructions replace the instructions
generated for encoding ordinary disjunctions.
 
As experiments in Section \ref{sec:lazyqueries} pointed out, the choice
of the actual lazy compilation variant does not matter with respect to 
the overhead introduced (except for lazy compilation per goal). We 
therefore chose to implement only the variant which compiles one 
complete disjunction at a time, as this made integration with the
existing query pack data structures easier. As explained in Section
\ref{sec:lazyqueries}, this means that each branch of a disjunction 
ends in a jump. In the implementation of the query packs execution,
this was already the case, so there are no extra emulator cycles in JIT
compiled code compared to the other compilation schemes. The memory
management aspects of the implementation will be discussed in 
Section~\ref{sec:memory}.

\subsection{Evaluation}

\begin{table}
\include{ace_experiments}
\caption{Experiments for query packs from a real world application.}
\label{tab:exppacks}
\end{table}

We evaluate (lazy) control flow compilation for query packs by 
running \textsc{Tilde}, but by letting it
generate query packs instead of conjunctions (as was the case for
Table~\ref{tab:aceconj}).  The experiments are performed on the ILP datasets
from Table~\ref{tab:aceconj}. Additionally, the query pack execution
mechanism allows us to do experiments on larger datasets, such as the 
HIV dataset \citeA{DTP}.

The timings in Table \ref{tab:exppacks} are in seconds: for compile \& run
and control flow, we give the sum of the total compilation time and the
total execution time; for lazy control flow compilation, no distinction can
be made, and so the total time for compilation and execution is given.
Additionally, we give for each dataset the share of query goals that are
never reached by the query execution.  Comparing the timings for the query
packs with the timings for the sets of queries in Table \ref{tab:aceconj} we
see that the query packs are considerably faster.

First, we compare control flow compilation with compile \& run.  For query
packs, control flow compilation is also up to an order of magnitude faster
than traditional compilation, even though the hipP system already has a
compiler that is optimized for dealing with large disjunctions
\cite{Vandecasteele2000} (in particular for the classification of variables
in query packs).  The execution times show the same characteristics as in
the experiments with the conjunctions in Table \ref{tab:aceconj}: control
flow has a faster execution than classical compilation.  For the ILP
application, the total time must be considered: the total time of control
flow is up to a factor 3 faster than compile \& run.
Note that this factor is higher for the query packs than for
the conjunctions.  The timings show that, for our benchmarks, the compilation time in 
compile \& run is systematically larger than the execution time for
all the examples such that the impact of improving the compilation has
a larger effect on the total times.

Table \ref{tab:lazy_disj} shows that lazy compilation has some
overhead, but we hoped that it would be compensated by avoiding the
compilation of failing parts in the query packs. This is indeed the case for
all datasets. As expected, the time gained by not compiling unused parts of
queries corresponds roughly with the measured amount of unreached goals.

The timings indicate that lazy
control flow compilation is the best approach for query packs.


\section{Memory Management Considerations} \label{sec:memory}

Experiments pointed out that the layout of the code and data in memory can 
have a big impact on the execution time of the queries.

Because the execution of the control flow compiled code needs to fetch the
data for its calls from the heap, the compiled code should be as close as
possible to the data it consumes to have good locality of data. We achieve
this by allocating control flow compiled code on the heap, and extending the
heap garbage collector to support this new data structure. Because of the
dynamic nature of lazy compiled code, control flow blocks can be scattered
across the heap during execution; the heap garbage collector moves these
blocks closer to each other during collection, which improves locality. 
Since queries have a volatile nature, the heap collector will also often
remove dead code blocks, which typically belong to old queries. Finally,
the code itself contains pointers to terms on the heap, which are
handled by the heap garbage collector as well.

\begin{table}
\include{fig_nocopy}
\caption{Impact of locality on execution times} \label{fig:nocopy}
\end{table}
Finally, the locality of the query goals themselves also has an impact on
execution time. During the query generation phase, other data is allocated
on the heap, which can lead to a situation where the goal terms of the query
(which will be used during execution of the control flow compiled code) are
scattered across the heap.  In all our experiments, the (possibly scattered)
term representing the complete query was copied before compiling it. This
ensured that all the terms used in the compiled code are allocated together
on the heap. The impact of leaving out the copying step is illustrated in
Table~\ref{fig:nocopy}. These are the results of running \textsc{Tilde} on
the same datasets as in Table~\ref{tab:exppacks}. Without copying the goals,
the execution time of control flow compiled code becomes slower than code
executed using the classical approach.  Although copying the query before 
compilation
costs some time, it improves the locality during the compilation step
itself.  The effect on the benchmarks is that it sometimes introduces a
slight overhead in some of the smaller benchmarks, but this is compensated
by the gain in execution time.  Control flow compilation with copy turns out
to be the best approach.

\section{Conclusions} \label{sec:conclusion} 
This paper presents a new method for faster compilation and execution
of dynamically generated queries: control flow compilation is up to an
order of magnitude faster than classical compilation, without affecting
the execution time. 
The benefits of control flow compilation versus classical compilation are
clear and are confirmed in the context of real world applications from the
ILP community. Moreover, the lazy variant provides additional speedup in the 
total time by not compiling unreached parts of the query. \\

Traditionally, Prolog implementations have implemented a form of JIT,
where compilation to WAM code or machine code happens at consult
time. Yap \cite{Yap} goes one step further and compiles a predicate to
abstract machine code at the first call to that predicate. BinProlog
\cite{BinProlog} and hipP switch back and forth between a compiled and
an interpreted form of dynamic predicates, based on the relative
frequency of modification and execution of the predicate. The
granularity of these JIT compilation forms is always a predicate,
while control flow compiled code can have a finer grained granularity
up to a literal. However, control flow compilation cannot be used for
compiling recursive predicates. 

Yap \cite{Yap}, which is used by the Aleph ILP system~\cite{Aleph}, has recently
introduced other implementation techniques for speeding up the
evaluation of many queries against many examples. In particular
tabling and dynamic indexing can speed up the query execution phase
considerably. Our control flow compilation schema is orthogonal to
these techniques and can be combined with them. Especially when
tabling is used, it is important to spend little time on compiling the
queries, as tabling avoids repeated execution of the same goal (or
prefix of a query). So, we expect that control flow compilation is
beneficial in combination with tabling. \\

Within the ILP setting, the applications of (lazy) control flow compilation
can be extended further. Firstly, we plan to adapt it to extensions of
query packs reported in \cite{Troncon2003b}. We expect control flow
compilation to yield the same speedups for these execution mechanisms as for
query packs.  However, the impact of laziness needs to be investigated.
In \cite{Ramon2004}, a technique for efficient theta-subsumption is
proposed which uses query pack execution.  It has to be investigated
whether lazy control flow compilation reduces the compilation time
enough in the particular setting that executes the query pack only
once, or that a pure meta-call based approach for the query packs performs 
better.

\section*{Acknowledgements}
We would like to thank the anonymous referees for their comments and
suggestions.


\bibliographystyle{acmtrans} 
\bibliography{cf.bib}


\end{document}

%% file: fig_cfvscr.tex
\begin{center}
\texttt{~~~query :- a(X,Y), ( b(Y,Z) ; c(Y,Z), d(Z,U); e(a,Y) ).}
\end{center}
\setlength{\tabcolsep}{.2cm}
\begin{center}
\texttt{
\begin{tabular}{l@{ }l|l}
    & \multicolumn{1}{c|}{\textrm{Control flow code}} 
        & \multicolumn{1}{c}{\textrm{Compiled code}} \\
 \cline{2-3}   
    & allocate 2            & allocate 4        \\
    &                       & bldtvar A1        \\
    &                       & putpvar Y2 A2     \\
    & cf\_call \&a(X,Y)     & call a/2          \\
    & trymeorelse L1        & trymeorelse L1    \\
    &                       & putpval Y2 A1     \\
    &                       & bldtvar A2        \\
    & cf\_deallex \&b(Y,Z)  & deallex b/2       \\
L1: & retrymeorelse L2      & retrymeorelse L2  \\
    &                       & putpval Y2 A1       \\
    &                       & putpvar Y3 A2       \\
    & cf\_call \&c(Y,Z)     & call c/2          \\
    &                       & putpval Y3 A1     \\
    &                       & bldtvar A2        \\
    & cf\_deallex \&d(Z,U)  & deallex d/2       \\
L2: & trustmeorelsefail     & trustmeorelsefail \\
    &                       & putpval Y2 A2     \\
    &                       & put\_atom A1 a    \\
    & cf\_deallex \&e(a,Y)  & deallex e/2      
\end{tabular}} 
\end{center}

%% file: fig_builtins.tex
\begin{center}
\texttt{
\begin{tabular}{l|l@{\hspace{2mm}}|l@{\hspace{2mm}}|l}
\hline \hline
\multicolumn{1}{c@{\hspace{5mm}}|}{\textrm{\textbf{Compiled code}}}
& \multicolumn{3}{c}{\textrm{\textbf{Control Flow code}}} \\
\multicolumn{1}{c|}{}
 & \multicolumn{1}{c}{\textrm{No Inlining}}  
 & \multicolumn{1}{c}{\textrm{Built-ins}}  
 & \multicolumn{1}{c}{\textrm{Special Built-ins}}  \\
\hline
\ldots              & \ldots              & \ldots & \ldots \\
call a/2            & cf\_call \&a(X,Y)   & cf\_call \&a(X,Y) & cf\_call \&a(X,Y) \\
putpval Y2 A1       &                     & putarg \&X A1      & \\
putpval Y3 A2       &                     & putarg \&Y A2      & \\
b\_smaller A1 A2    & cf\_call \&(X<Y)    & b\_smaller A1 A2   & cf\_smaller \&X \&Y\\
\ldots              & \ldots              & \ldots & \ldots \\  
\hline \hline
\end{tabular}}
\end{center}
\vspace{-1\baselineskip}

%% file: tab_artif.tex
\begin{minipage}{\textwidth}
\begin{tabular}{cccclrr}
\hline \hline
\multicolumn{4}{c}{\textbf{Query}} 
& \textbf{Experiment} & \multicolumn{2}{c}{\textbf{Time}} \\
G\footnote{Number of goals in a branch} 
& B\footnote{Branching factor of each disjunction}
& D\footnote{Nesting depth of disjunctions}
& T\footnote{Total number of goals ($ = G\sum_{n=0}^{n=D}B^n$)} 
&
& Comp.\footnote{Compilation time of the query (in ms.)} 
& Exec.\footnote{Execution time of the query (in ms.)} \\
\hline
 5 &  5 & 4 & 3905
              & Meta-call      & \na   & 2.14 \\
        & & & & Compile \& Run & 288.0 & 0.384 \\
        & & & & Control Flow   & 30.8 & 0.198 \\
10 &  5 & 4 & 7810
              & Meta-call      & \na   & 4.196 \\
        & & & & Compile \& Run & 668.0 & 0.734 \\
        & & & & Control Flow   & 62.6  & 0.36 \\
 5 & 10 & 4 & 55555
              & Meta-call      & \na    & 33.676 \\
        & & & & Compile \& Run & 6368.0 & 5.64 \\
        & & & & Control Flow   & 457.4  & 3.15 \\
10 & 10 & 4 & 111110
              & Meta-call      & \na     & 64.94 \\
        & & & & Compile \& Run & 13876.0 & 10.812 \\
        & & & & Control Flow   & 847.8   & 5.718 \\
 5 &  5 & 6 & 19531
              & Meta-call      & \na     & 59.198 \\
        & & & & Compile \& Run & 11596.0 & 9.93  \\
        & & & & Control Flow   & 758.0   & 5.402 \\
\hline \hline
\end{tabular}
\vspace{-2\baselineskip}
\end{minipage}

%% file: tab_aceconj.tex
\begin{minipage}{\textwidth}
\begin{tabular}{lrrlrrr}
\hline \hline
Dataset   & Queries\footnote{Total number of different queries executed} &
Runs\footnote{Average number of examples on which a query is run}  & Experiment        & Comp.\footnote{Total compilation time of all
queries (in seconds)} & Exec.\footnote{Total execution time of all queries (in
seconds)} & Total \\
\hline
\textsf{Mutagenesis}    & 2021  &  69.51 
                            & Meta-call       & \na  & 1.43 & 1.43\\
                        & & & Compile \& Run & 1.30 & 1.06 & 2.36 \\
                        & & & Control Flow   & 0.21 & 1.02 & 1.23 \\
\textsf{Bongard}        & 9335  & 244.77 
                            & Meta-call       & \na  & 24.70 & 24.70 \\
                        & & & Compile \& Run & 4.98 & 21.34 & 26.32 \\
                        & & & Control Flow   & 0.91 & 21.18 & 22.09 \\
\textsf{Carcinogenesis} & 48399 & 103.07 
                            & Meta-call       & \na   & 108.81 & 108.81 \\
                        & & & Compile \& Run & 17.50 &  65.30 &  82.80 \\
                        & & & Control Flow   &  2.24 &  59.51 &  61.75 \\
\hline \hline
\end{tabular}
\vspace{-2\baselineskip}
\end{minipage}

%% file: tab_lazyartif.tex
\begin{minipage}{\textwidth}
\begin{tabular}{ccclr}
\hline \hline
\multicolumn{3}{c}{\textbf{Query}} 
& \textbf{Experiment} 
& \multicolumn{1}{c}{\textbf{Time}} \\
G\footnote{Number of goals in a branch} 
& B\footnote{Branching factor of each disjunction}
& D\footnote{Nesting depth of disjunctions} 
&
& Total\footnote{Total Compilation + Execution time
of the query (in ms.)} \\
\hline
 5 &  5 & 4 
            & Per Goal & 55 \\
        & & & Per Conjunction & 34 \\
        & & & Per Disjunction & 32 \\
        & & & No Laziness & 28 \\
10 &  5 & 4 
            & Per Goal & 111 \\
        & & & Per Conjunction & 60 \\
        & & & Per Disjunction & 59 \\
        & & & No Laziness & 59 \\
\hline \hline
\end{tabular}
\vspace{-2\baselineskip}
\end{minipage}

%% file: ace_experiments.tex
\begin{minipage}{\textwidth}
\begin{tabular}{lclrrr}
\hline \hline
Dataset     & Unused\footnote{Total \% of the query code
that is never executed} & Experiment & Comp.\footnote{Total compilation time of all
queries (in seconds)} & Exec.\footnote{Total execution time of all queries (in
seconds)} & Total\footnote{Total query evaluation time (= Comp. + Exec.)} \\
\hline
\textsf{Mutagenesis}    & 17\% & Compile \& Run    & 0.52 & 0.11 & 0.63 \\
                        &      & Control Flow      & 0.07 & 0.10 & 0.17 \\
                        &      & Lazy Control Flow & \na & \na    & 0.14 \\
\textsf{Bongard}        & 51\% & Compile \& Run    & 1.91 & 1.17 & 3.08 \\
                        &      & Control Flow      & 0.28 & 1.15 & 1.43 \\
                        &      & Lazy Control Flow & \na & \na & 1.37 \\
\textsf{Carcinogenesis} & 32\% & Compile \& Run    & 7.39 & 4.63 & 12.02 \\
                        &      & Control Flow      & 0.81 & 3.81 & 4.62 \\
                        &      & Lazy Control Flow & \na & \na    & 4.34 \\
\textsf{HIV}            & 74\% & Compile \& Run    & 209.47 & 191.68 & 401.15 \\
                        &      & Control Flow      & 27.13 & 178.53 & 205.66 \\
                        &      & Lazy Control Flow & \na & \na & 186.22\\
\hline \hline
\end{tabular}
\vspace{-2\baselineskip}
\end{minipage}

%% file: fig_nocopy.tex
\begin{minipage}{\textwidth}
\begin{tabular}{llrrr}
\hline \hline
Dataset     & Experiment        & Comp.\footnote{Total compilation time of all
queries, including the time to copy the query (in seconds)} & Exec.\footnote{Total execution time of all queries (in
seconds)} & Total\footnote{Total query evaluation time (= Comp. + Exec.)} \\
\hline
\textsf{Mutagenesis} 
    & Compile \& Run         & 0.52 & 0.11 & 0.63 \\
    & Control Flow (No Copy) & 0.08 & 0.10 & 0.18 \\
    & Control Flow           & 0.07 & 0.10 & 0.17 \\
\textsf{Bongard} 
    & Compile \& Run         & 1.91 & 1.17 & 3.08 \\
    & Control Flow (No Copy) & 0.33 & 1.19 & 1.52 \\
    & Control Flow           & 0.28 & 1.15 & 1.43 \\
\textsf{Carcinogenesis} 
    & Compile \& Run         & 7.39 & 4.63 & 12.02 \\
    & Control Flow (No Copy) & 0.70 & 4.11 &  4.81 \\
    & Control Flow           & 0.81 & 3.81 &  4.62 \\
\textsf{HIV} 
    & Compile \& Run         & 209.47 & 191.68 & 401.15 \\
    & Control Flow (No Copy) & 27.81 & 193.90 & 221.71 \\
    & Control Flow           & 27.13 & 178.53 & 205.66 \\
\hline \hline
\end{tabular}
\vspace{-2\baselineskip}
\end{minipage}